\documentclass[10pt,twocolumn]{article} 
\usepackage{simpleConference}
\usepackage{times}
\usepackage{graphicx}
\usepackage{amssymb}
\usepackage{url,hyperref}
\usepackage{mathtools}  

\usepackage{stackengine,graphicx}
\stackMath
\newlength\matfield
\newlength\tmplength
\def\matscale{1.}
\newcommand\dimbox[3]{%
  \setlength\matfield{\matscale\baselineskip}%
  \setbox0=\hbox{\vphantom{X}\smash{#3}}%
  \setlength{\tmplength}{#1\matfield-\ht0-\dp0}%
  \fboxrule=1pt\fboxsep=-\fboxrule\relax%
  \fbox{\makebox[#2\matfield]{\addstackgap[.5\tmplength]{\box0}}}}
\newcommand\raiserows[2]{%
   \setlength\matfield{\matscale\baselineskip}%
   \raisebox{#1\matfield}{#2}}
\newcommand\matbox[5]{
  \raisebox{2.3ex}{\rotatebox[origin=center]{90}{\scalebox{.45}{#2}}}\,%
  \stackon{\dimbox{#1}{#3}{#5}}{\scalebox{.45}{#4}}}
\newcommand\titlebox[3][30ex]{\stackanchor{#2}{\scalebox{.45}{\parbox{#1}{\centering #3}}}}

\begin{document}

\title{Audio Spectrogram Factorization for Classification \\
       of Telephony Signals below the Auditory Threshold}
\author{Iroro Orife, Shane Walker and Jason Flaks\\
Marchex Inc., 520 Pike, Seattle, WA, 98101 \\
}

\maketitle
\thispagestyle{empty}

\begin{abstract}
   Traffic Pumping attacks are a form of high-volume SPAM that target telephone networks, defraud customers and squander telephony resources. One type of call in these attacks is characterized by very low-amplitude signal levels, notably below the auditory threshold. We propose a technique to classify so-called ``dead air" or ``silent" SPAM calls based on features derived from factorizing the caller audio spectrogram. We describe the algorithms for feature extraction and  classification as well as our data collection methods and production performance on millions of calls per week. 
   
\end{abstract}

\section{Introduction}
The Public Switched Telephone Network (PSTN) is a collection of interconnected telephone networks that abide by the International Telecommunication Union (ITU) standards allowing telephones to intercommunicate. In recent decades, telephone service providers have modernized their network infrastructure to Voice over Internet Protocol (VoIP) to take advantage of the availability of equipment, lower operating costs and increased capacity to provide voice communication and multimedia sessions. These advances also make it easy to automate the distribution of unsolicited and infelicitous call traffic colorfully known as robocalling or \textbf{SP}am over \textbf{I}nternet \textbf{T}elephony (SPIT) \cite{VoIP_spam, rosenberg2008session}. For a comprehensive background on the ecosystem, pervasiveness of telephone SPAM, the various actors and the impact to consumers, refer to \cite{tu2016sok, Phone_fraud}.

Traffic Pumping (TP) is one class of robocall that starts when small, usually rural, Local Exchange Carriers (LXC) partner with telephone service providers (typically with toll-free numbers) to route the latter's calls. The LXC flood their own telephone networks with call traffic to boost their call volume and the inter-carrier revenue-sharing fees they are owed by long-distance, cross-regional Interexchange Carriers (IXC), per the Telecommunications Act of 1996 \cite{WikiTrafficPumping, FCCTrafficPumping}. Confusing matters further, in the call-routing chain each carrier only sees the preceding and following carrier. Therefore, no single player has the full route back to the spammer, which complicates provenance tracking and completely stopping TP \cite{dialogtech}.

The telephony stack behind the Marchex call and speech analytics business handles over one million calls per business day, or decades of encrypted audio recordings per week \cite{walker2017semi}. This volume makes our telephony stack particularly susceptible to attempts to anonymously establish automated voice sessions. TP caller audio has a number of different qualitative characteristics, it may be recordings (e.g. a section of an audio-book or music), broadband noise or signaling tones (e.g. busy, ring, fax, modem). The caller audio may also simply be ``dead air", where no one answers and there is no audible sound. By this, we do not suggest \emph{digital silence}, but very low level signal levels below the threshold of audibility. Developing countermeasures for this latter category of ``silent" signal is focus of this report. 

Effectively mitigating TP is an adversarial match with the spammer. As countermeasures are deployed, spammers are motivated to find new means to dynamically evade. Inbound calls also require immediate attention, which at scale introduces real-time constraints on SPAM classifiers. We contrast this with email SPAM mitigation which can often be deferred for future analysis offline. Finally, our call based business requires very high accuracy, as false positives, i.e. mistakenly blocked calls, are detrimental to our customers \cite{tu2016sok}.
\\
\\
The main contributions of this paper are as follows:
\begin{description}
  \item[$\bullet$]  We propose a novel countermeasure for ``dead air" VoIP SPAM using features derived from the Singular Value Decomposition (SVD) of the audio spectrogram.
  \item[$\bullet$]  Given the first two seconds of a call, we demonstrate the efficacy of a Random Forest classifier trained on these features to classify SPAM at production scale.
\end{description}
This paper is organized as follows, Section 2 briefly summarizes related work mitigating SPAM. Section 3 details the algorithms for audio feature extraction. Section 4 describes our use of Random Forests for classification. Section 5 and 6 detail our experiments and results, while Section 7 discuss our post-attack experiments. Section 8 concludes with ideas for future work.

\section{Related Work}

VoIP SPAM countermeasures broadly fall into two categories. The first involves the use of Call Request Header (CRH) metadata e.g. Caller ID or caller's terminal device, in concert with white and black lists. Previous behavior and reputation systems also use Caller ID to track behavior and reputation scores. However CRH metadata is not always present nor reliably propagated while Caller ID numbers can be easily and cheaply falsified with open-source VoIP PBX systems such as Asterix PBX or FreeSWITCH \cite{Caller_ID_spoofing, venkatasubbareddyimplementation}. 
 
The second approach involves training machine-learned models to classify calls based on telephony control signal features or the call audio itself. When the call audio is based on a recording, landmark-based audio fingerprinting (e.g. music-id app Shazam) is commonly used for identification \cite{YapHash2012, wang2003industrial}. Acoustic pattern analysis features include voice-codec signatures, \{VoIP vs.\ PSTN\} packet loss patterns or channel noise-profiles \cite{Balasubramaniyan2010PinDr0pUS, tu2016sok}. These systems are often coupled with interactive challenge-response tests (e.g. Audio-CAPTCHAs or Interactive voice response (IVR) ``Turing tests") to further verify the identity of suspicious callers and minimize false positives.

For our specific task with ``dead air" audio, we know of only one related effort, which focused on mobile device identification using Mel-frequency cepstral coefficients (MFCC) features. To account for the absence of an audible speech signal, Jahanirad et al. computed the entropy of the Mel-cepstrum, discovering that sections of ``silent" signal resulted in high-valued entropy-MFCC features which were effective in discriminating between different mobile device models \cite{jahanirad2014blind}. This work suggests there are characteristic features of a source device or the transmission channel even in the absence of a speech signal.

\section{Feature Extraction}
In telephony audio, background noise levels can have a fair amount of energy, so simple energy heuristics do not effectively identify ``silence". This observation informed investigations beyond simple activity detection models. The mission was now to find discriminable features given the first two seconds of a ``silent" call. This short duration was chosen because it can fit within the cadence of an extra ringing tone. Analyses of the audio revealed no dominant harmonicity, fundamental frequencies nor reliable temporal envelope descriptions. Zero crossing and spectral shape statistics required a minimum signal level above the noise-floor to be of any use as an input to a machine learning classifier \cite{jahanirad2014blind}. In other words, peak levels at -50 dBFS and average levels around -72 dbFS in our 16-bit system corresponded to a very low dynamic range and a minimum signal-to-noise ratio, which limited the signal analysis options.

\subsection{Spectrogram Representation}
Spectrograms are two dimensional representations of sequences of frequency spectra with time along the horizontal axis and frequency along the vertical. The color and/or brightness illustrates the magnitude of the frequency $k$ at time-frame $n$. A spectrogram compactly addresses the need to represent a \emph{sequence} of audio features over the two second duration required for the task. Empirically, they are a practical choice, possessing more information that any single audio feature investigated above, but with lower dimensionality than the time domain waveform \cite{wyse2017audio}.

To compute the spectrogram, we decode the first two seconds of mono 8kHz $\mu$-law audio from our VoIP PBX to 16-bit PCM and then use a Short Time Fourier Transform (STFT) to compute the complex-valued spectrogram $S_c(n,k)$ with $n$ and $k$ as the time-frame and frequency bin indices respectively: 
\begin{gather} 
 S_c(n,k) = \displaystyle\sum_{l=-\frac{W}{2}}^{\frac{W}{2} - 1} w(l) \cdot x(l + nh) \cdot e^{-2\pi ilk/W}
\end{gather} 
To compute the STFT, the audio signal $x(n)$ is sliced into overlapping \emph{segments} of equal length, $W$ samples wide. Each segment is offset in time by a hop size value $h$. A Hann or raised-cosine window $w(l)$ is multiplied element-wise with each segment. This ``windowing" acts a tapering function to reduce spectral leakage during Fast Fourier Transforms (FFT) \cite{lyon2009discrete}. Finally, an FFT of size $W$ is computed separately on each windowed waveform segment to generate a complex spectrogram \cite{scipy.signal.hanning}. The magnitude spectrogram $X(n,k)$ is the complex modulus of $S_c(n,k)$. 
\begin{gather} 
 X(n,k) = |S_c(n,k)| 
\end{gather} 
For a complex number $z$, its complex modulus is defined as $|z| = {\sqrt {x^{2}+y^{2}}}$, where $x$ and $y$ denote the real and imaginary parts respectively. 
We compute the magnitude spectrogram rather than the power or mel-spectrogram, as we observed that the dynamic scale of the magnitude representation better suited to low-intensity signals \cite{Virtanen}.

\subsection{Spectrogram Decomposition} \label{SpectrogramDecomposition}
Spectrogram factorization has been widely used for its flexibility modeling compositional mixtures of sounds from disparate sources. Applications include source separation \cite{smaragdis2007supervised}, music information retrieval, environmental sound classification \cite{wang2005musical}, instrument timbre classification \cite{kaminiarz2007mpeg}, drum transcription \cite{orife2001riddim}, de-reverberation \cite{jukic2015multi} and noise-robust speaker recognition \cite{hurmalainen2015noise}.  

Because we are strictly looking to classify, not recover entire components, we conceptualize the ``silent" audio magnitude spectrogram as an additive mixture of spectral basis vectors corresponding to various source and transmission channel factors. We represent constituent elements thusly, in both frequency and time, assuming that only the scale of each spectral basis is time-variant \cite{sakata2016applied}. 

There are different decompositions of a matrix based on its properties, e.g. square vs.\ rectangular, symmetric, non negative elements or positive eigenvalues. In applications where mixtures of sounds sources in multicomponent signals are modeled by their distribution of time/frequency energy, latent sources are recovered using Principal component analysis (PCA). This is accomplished by a singular value or eigen-decomposition of the autocorrelation matrix, $XX^T$ of the spectrogram $X$ \cite{orife2001riddim}. 

On the magnitude spectrogram, we use the Singular Value Decomposition (SVD) because it yields a ``deeper" factorization and produces unique factors. Additionally, singular values are useful in understanding the most important spectral bases \cite{CBoling15}. The SVD of an $n$$\times$$m$ matrix $X$ is the factorization of $X$ into the product of three matrices: 
\begin{gather} 
X = UDV^T
\end{gather} 
\[\renewcommand\matscale{.75}
\matbox{7}{frequency}{8}{time}{%
  \titlebox[20ex]{X\\}{Spectrogram}}
= 
\matbox{7}{frequency}{3}{dimensions}{\titlebox[10ex]{U}{basis spectra}} 
\raiserows{2}{\matbox{3}{dimensions}{3}{dimensions}{\titlebox{D}{weights}}} 
\stackunder[6pt]{%
  \raiserows{2}{\matbox{3}{dimensions}{8}{time}{\titlebox{V^T}{time activations}}}%
}{\scalebox{.6}{}}
\] \\
where the columns of the $n$$\times$$d$ matrix $U$ and the $m$$\times$$d$ matrix $V$ consist of the left and right singular vectors, respectively. $D$ is a $d$$\times$$d$ diagonal matrix whose diagonal entries are the singular values of $X$, representing the ``strength" of each spectral basis.
 
 \begin{figure}[h]
  \begin{center}
    \includegraphics[width=3.3in]{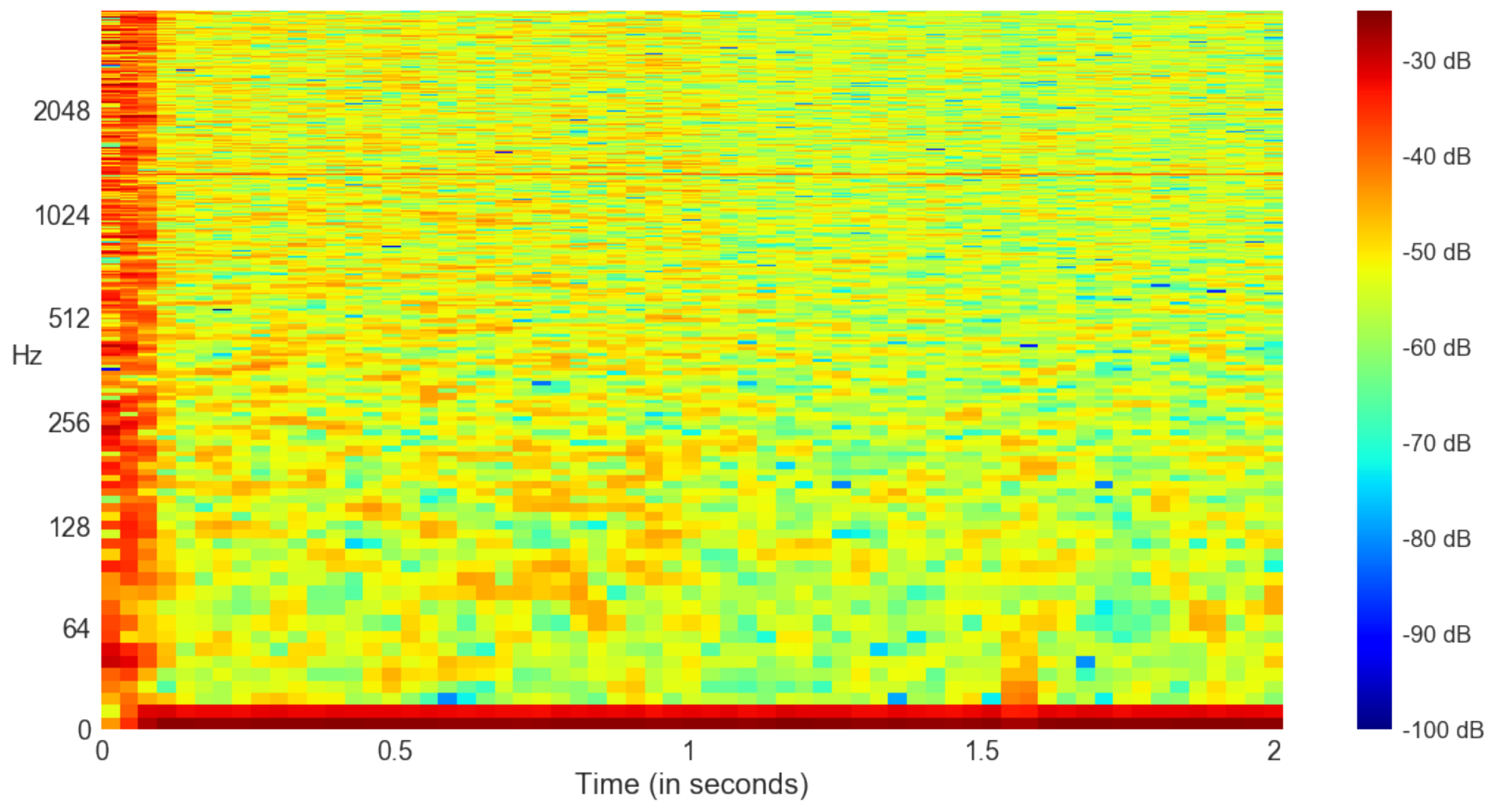}
  \end{center}
  \caption{\small Log-frequency Power Spectrogram of first two seconds of a TP ``silent" call. Power Spectrogram plotted for visualization}
\end{figure}

\section{Classification}

To classify spectral bases feature vectors from $U$, we trained a Random Forest (RF) classifier. The latter is an ensemble learning method for supervised classification tasks. Ensembles are a divide-and-conquer approach where a group of ``weak learners" band together to form a ``stronger learner". RFs arise from a machine learning modeling technique called a decision tree, i.e. our weak learner. For classification, decision trees make predictions based on observations about the feature data, represented by the branches, to judgements about the data's class, represented by the leaves.

RFs work by constructing a multitude of random decision trees during training, yielding the class that is the \emph{mode} of all the classes. To classify new objects from an input feature vector, the latter is processed by each tree, which gives a classification or ``vote" for a particular class. The forest chooses the classification with the most votes among all the trees in the forest. 

\begin{figure}[!htbp]
  \begin{center}
    \includegraphics[width=3.3in]{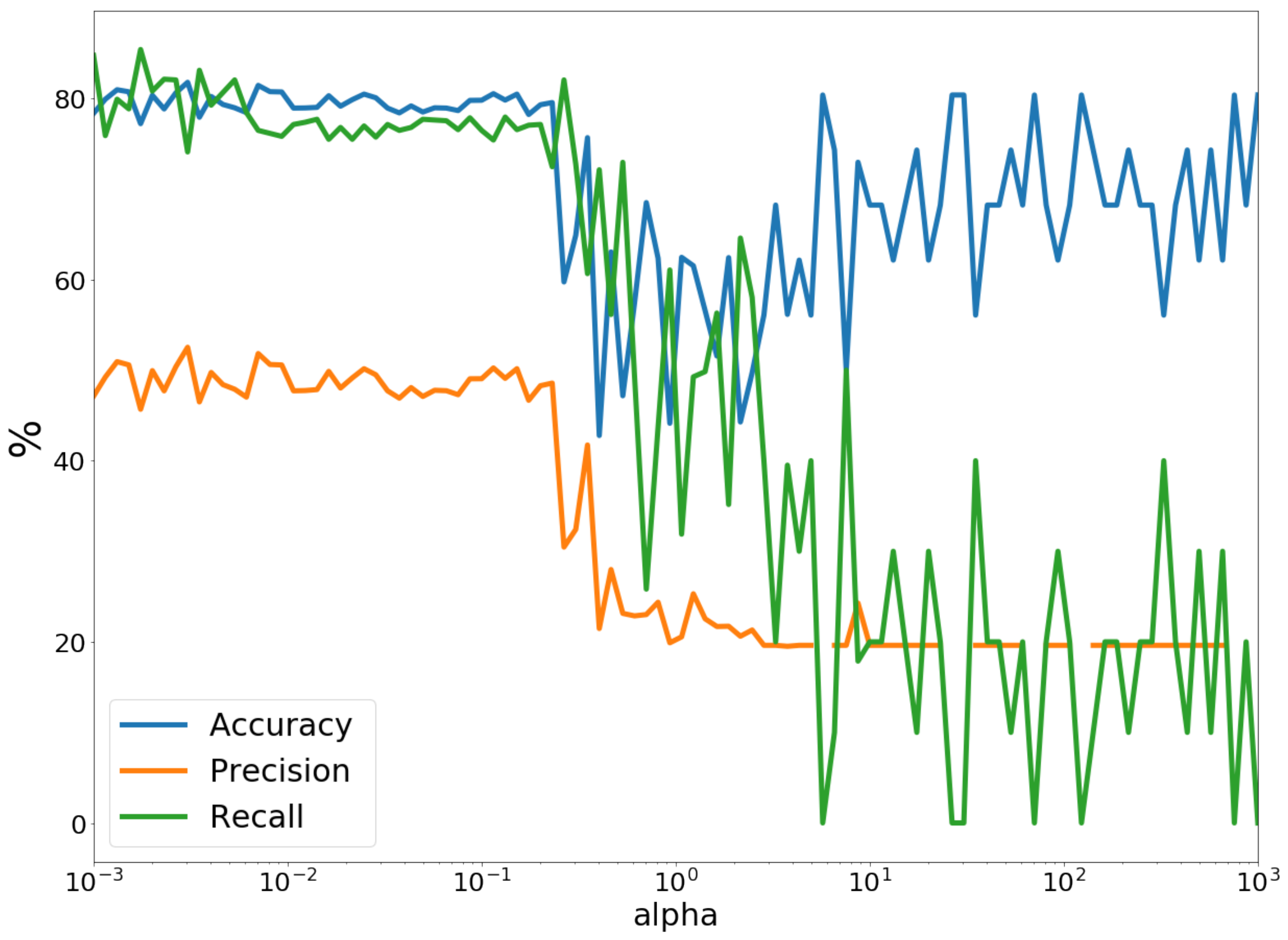}
  \end{center}
  \caption{\small Stratified shuffle split cross-validation results over the regularization term \emph{alpha}, for \texttt{SGDClassifier}, an SVM classifier with SGD training}
  \label{fig:xvalid-sgd}
\end{figure}

\begin{figure}[t]
  \begin{center}
    \includegraphics[width=3.3in]{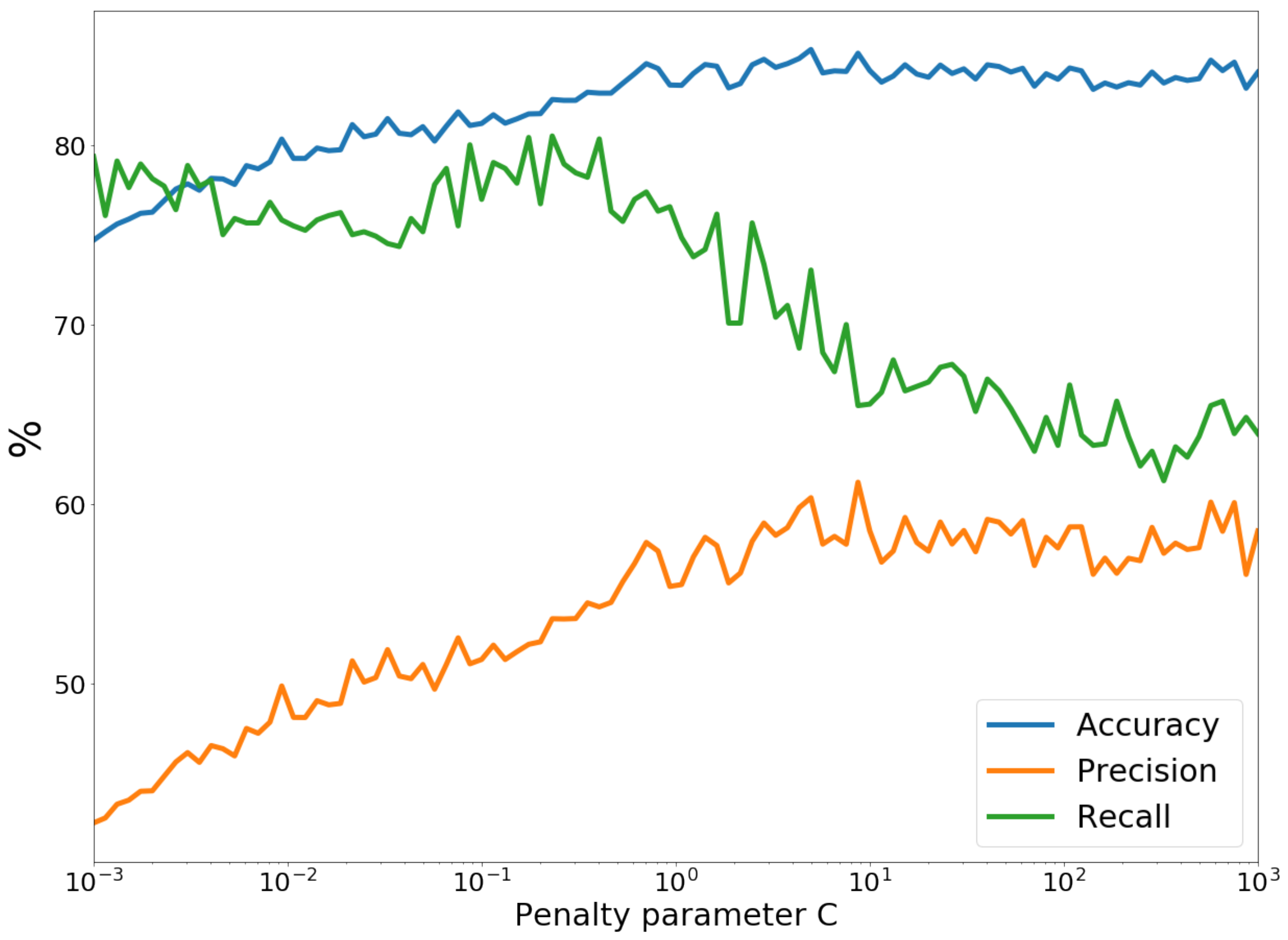}
  \end{center}
  \caption{\small Stratified shuffle split cross-validation results over the penalty parameter \emph{C}, for a \texttt{LinearSVC}}
  \label{fig:xvalid-svc}
\end{figure}

RFs are a good choice for our task because of superb accuracy on datasets of different sizes. They also less prone to balance error in the case that there are imbalances in the number of training examples for each class. In our case we have a lot more HAM\footnote{HAM are desirable calls, in contrast to SPAM} than SPAM. We also considered how well they generalize (i.e. do not overfit) and their computational performance during prediction \cite{RandomForests}. In contrast to linear support vector machine classifiers, see Figures \ref{fig:xvalid-rf}-\ref{fig:xvalid-svc}, with RFs we were able to obtain good precision and business acceptable recall.

\section{Experiments}

During a TP attack, an extra 10,000 to 33,000 ``silent" calls per day may be  handled by the telephony stack. Business requirements specified that two seconds was the maximum permissible latency for this category of SPAM analysis before bridging calls. To label calls:

\begin{description}
  \item[$\bullet$] We enabled the call processors to write the first two seconds of audio data to disk during the collection period. Each audio file was named with a unique call-id.
  \item[$\bullet$] We used a metric called Caller Speech (CS) emitted by the Voice Activity Detector (VAD) in our downstream ASR system \cite{walker2017semi} to find call-ids with zero Caller Speech.
  \end{description}

This latter metric tracks the number of times the VAD activates to capture caller utterances. For ``dead air" calls, this value was always zero. Because we compute CS \emph{after} the call has ended, this technique was only useful to select calls for labeling training data.

We collected 8,000 calls during one day during the attack, filtering calls that failed the IVR Turing test in concert with zero CS fields, we were able to quickly and definitively label 256 ``silent" SPAM calls. Another 692 failed the IVR test and had null CS values, but exhibited the same ``dead" air features upon manual inspection. We also labeled 1,500 calls with positive CS values that successfully passed the IVR test.

\begin{figure}[h]
  \begin{center}
    \includegraphics[width=3.3in]{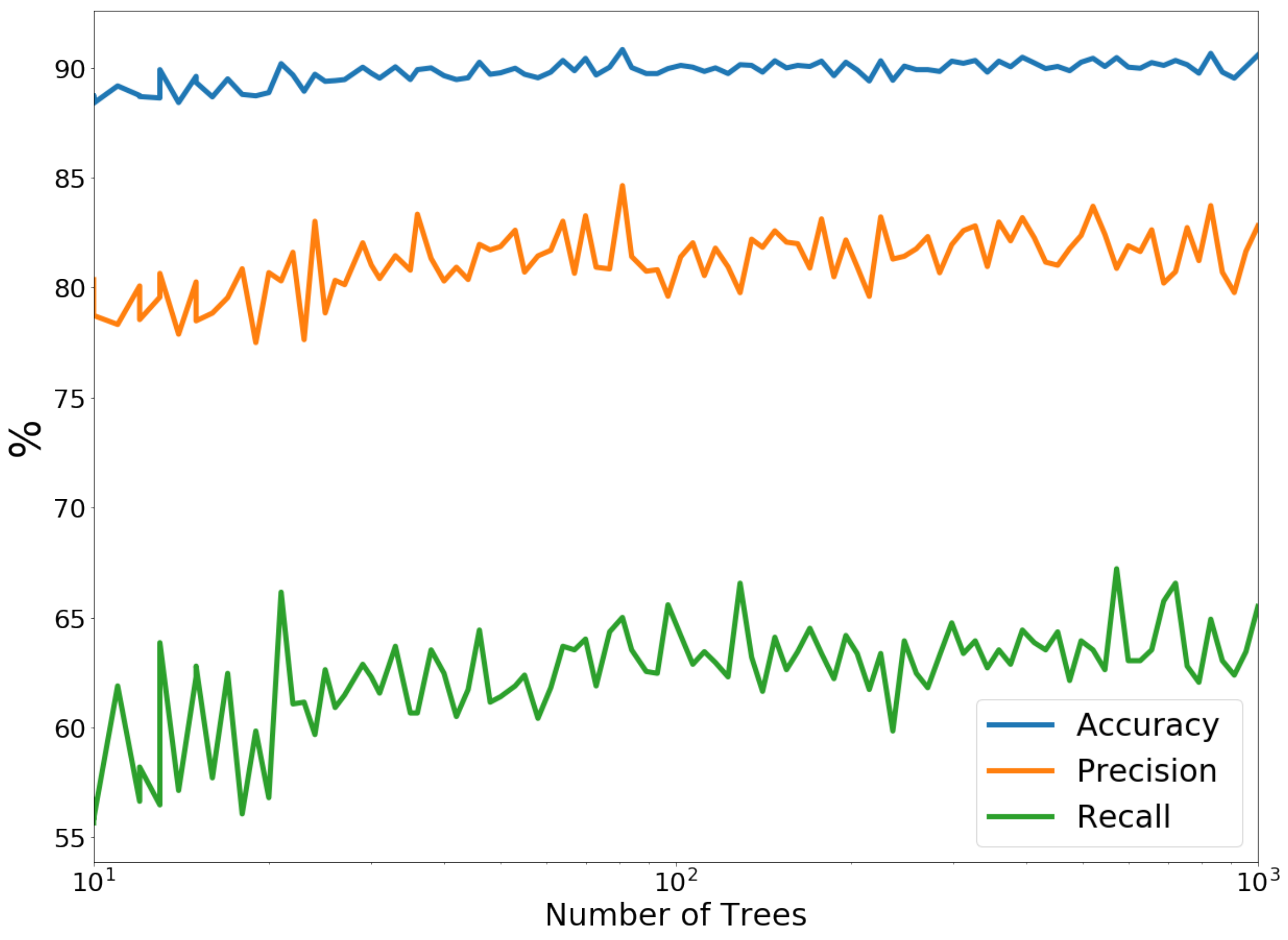}
  \end{center}
  \caption{\small Stratified shuffle split cross-validation results over the number of trees in the forest for \texttt{RandomForestClassifier}}
  \label{fig:xvalid-rf}
\end{figure}

For feature extraction, we used the Python frameworks, \emph{numpy} and \emph{librosa} \cite{McFee2015librosaAA}. For classification, we used the \emph{scikit-learn} \texttt{RandomForestClassifier} \cite{scikit-learn}, trained for 100 iterations on the top 3 bases of $U$ (Section \ref{SpectrogramDecomposition}). For cross validation, we used a stratified shuffle split cross-validator over the total number of trees, i.e. the data is shuffled each time, before a new train/test split is generated. Refer to Figure \ref{fig:xvalid-rf}.

To establish a baseline performance with other classifiers, we experimented with two other linear models in \emph{scikit-learn}, see Table \ref{tab:xvalid-results}. The first is the Linear Support Vector Classification, \texttt{LinearSVC}, which uses a \texttt{liblinear} implementation that penalizes the intercept and minimizes the squared hinge loss. The second is \texttt{SGDClassifier} which optimizes the same cost function as \texttt{LinearSVC} but with the hinge loss and stochastic gradient descent in lieu of exact gradient descent. While both are linear kernel methods, the first is optimized to scale to a large number of samples, while the second supports minibatch training and may generalize better \cite{scikit-learn}.

\section{Initial Results}

Evaluating the system performance entails comparing the model's predicted value to ground truth. There are four possible outcomes of the system's performance: A True Positive (TP) is a correct SPAM prediction, a True Negative (TN) is a correct HAM prediction. A False Positive (FP) is an incorrect SPAM prediction, a False Negative (FN) is a incorrect HAM classification, i.e. an actual SPAM call was able to get through.

Precision, also known as positive predictive value, is the percentage of positive SPAM classifications that were actually SPAM. Intuitively, precision tells us how correct the classifier is when it predicts that a call is SPAM. Precision suffers as the number of FPs grow. Similarly, recall, also called sensitivity, tells us how many of the positive cases did the classifier get from the total number of positive cases. Recall suffers as the number of FNs grow. Accuracy is the proportion of correctly identified calls (TP + TN) to all calls.
\begin{gather} 
Precision = \frac{TP}{TP + FP} \\[5pt]
Recall = \frac{TP}{TP + FN}
\end{gather} 
From a high-volume business perspective, it is much less egregious to let in the occasional SPAM call (FN), than it is to label and reject legitimate calls as SPAM (FP). In other words, high precision (low FPs) is more important than high recall (low FNs).
\begin{table}[!htbp]
  \begin{center}
    \begin{tabular}{l|l|l|l} \hline
       Model & Precision & Recall & Accuracy\\
      \hline
      Linear SVC & 58.57 & 63.60 & 84.04  \\
      Linear SVC with SGD  & 81.85 & 52.56 & 76.31  \\
      \textbf{Random Forest} & \textbf{83.82} & \textbf{63.27} & \textbf{90.40} \\
      \hline
    \end{tabular}
  \end{center}
    \caption{\small Classifier performance}
    \label{tab:xvalid-results}
\end{table}

In examining the precision and recall figures from cross-validation, a model of 100 trees was chosen for production to foremost optimize precision (84\%) and overall accuracy (90\%). This practical choice was also influenced by our recognition of the limits and biases of our small and class imbalanced dataset.

As the realtime and low-latency requirements for classification across thousands of concurrent calls made any kind of manual verification very impractical, to further minimize hanging up on real customers (FPs), we introduce the minor inconvenience of an IVR Turing test for all positive SPAM predictions, requiring callers to press a randomly chosen numeric key to proceed. Calls that do not make it past the IVR are labeled as \texttt{REJECTED\_CALLER\_SILENCE} in the production call-log.

\section{Post-Attack Experiments}

The nature of TP attacks is such that they are high intensity over a short period of time, from days to weeks, after which SPAM levels subside, returning to a ``steady-state" or low-grade level. This reduction could be externally motivated or in response to our blocking efforts. After this ``dead air" attack abated, to re-evaluate the performance of the classifier and understand any new characteristics of the low-prevalence SPAM, we captured a full day of audio from calls on a single production host, some 69,900 calls.
\begin{table}[!htbp]
  \begin{center}
    \begin{tabular}{l|l} \hline
       Spam Type & Count \\
      \hline
      	CALLER\_FAX &  611 \\
		\textbf{CALLER\_SILENCE} & \textbf{233} \\
		CALLER\_RECORDING & 87 \\
		CALLER\_NOISE &   32 \\ 
		CALLER\_BUSY  &  5 \\
      \hline
      Total  &  968 \\
    \end{tabular}
  \end{center}
  \caption {\small Prevalence of SPAM types in 968 post-attack calls captured on one host} 
      \label{tab:confirmed-spam-prevalence}
\end{table}

Of these, 3,096 were marked as a kind of SPAM (not just ``silent") by the classifiers on that host. Cross-referencing these calls with call-stack metadata gave us 968 confirmed calls for which we had both audio and a metadata tag of SPAM that had been rejected (i.e. the call-stack hung up or the call failed IVR Turing test). Table \ref{tab:confirmed-spam-prevalence} lists the prevalence of different kinds of SPAM, with ``silent" SPAM in boldface. 

After matching 968 rejected calls, we were left with 2,128 calls which were \emph{not} rejected by the call-stack, all of which were classified by our model as ``silent" SPAM. This extremely high count of predicted ``silent" SPAM calls suggested a very high FP rate and that our error analysis was far from complete.

Tabulating the call result metadata for each of the 2,128 predicted ``silent" calls revealed a variety of outcomes, other than failing to pass the IVR Turing test. Most notable were calls which ended \emph{during} the IVR Turing test or were bridged only to be ended by the caller anyway. Only 363 were ended by the agent. In Table 3 we enumerate all other ways a predicted ``silent" call was ended. This error analysis elucidates the distribution of errors and notably what kinds of audio calls were FP.
\begin{table}[!htbp]
  \begin{center}
    \begin{tabular}{l|l|l} \hline
      Call Result & Ended By & Count \\  
      \hline
      BRIDGED & CALLER & 1159 \\
      IVR\_ABANDON & CALLER & 547 \\
	  BRIDGED & AGENT & 363 \\
	  BRIDGE\_ABANDON & CALLER & 37 \\ 
	  BRIDGE\_TIMEOUT & APP & 13 \\
	  NO\_ELIGIBLE\_AGENTS & APP & 4 \\  
	  BRIDGE\_TIMEOUT & CALLER & 2 \\ 
      BRIDGE\_ABANDON & AGENT & 1 \\
      NO\_ELIGIBLE\_AGENTS & CALLER & 1 \\
      BRIDGE\_OUT\_OF\_SERVICE & APP & 1 \\
      \hline
    \end{tabular}
  \end{center}
  \caption {\small Reasons 2,128 ``silent" calls were ended} 
  \label{tab:unjoined-reasons}
\end{table}

Most surprisingly, once we started programmatically analyzing audio predicted as ``silent" SPAM, regardless of whether it was rejected or abandoned by the agent/caller, 98.6\% \{2100/2128, 229/233\} of all predicted ``silent" calls were, in this post-attack test, pure digital silence! 

There are many conditions under which digital silence is correctly produced by the call-stack, for example at the beginning of HAM and some SPAM calls, during transfers or right before a ring tone. Our error analysis illuminated the fact that digital silence needed to have been better considered when labeling call segments. Had we recognized the prevalence and interplay with ``dead air" silence, we would have labeled calls with mixtures differently, at a finer resolution or factored out any digital silence from positive ``dead air" training segments.

This \emph{shift} in the distribution of audio features between the training and test phases versus running ``in the wild", over time,  raises questions about the steady-state efficacy of the classifier and the need to keep the model up to date in this adversarial match. These questions are common in machine learning practice, i.e. when an algorithm is trained in a lab setting without seeing the full breadth of data from the production system, carefully formulated questions must be asked and exhausted to ensure that models continue to perform satisfactorily.

\section{Future Work}

In addition to the practical machine learning matters and keeping the SPAM classifiers regularly updated, there are a number of interesting algorithmic avenues for future work. These include various neural network alternatives to non-negative matrix factorization (NMF) \cite{DBLP:journals/corr/SmaragdisV16} as well as fine-tuned, state of art image classification models \cite{finetuningkeras}. Fine-tuning is the process of a trained model's architecture and weights as a starting point, freezing lower layers and training on our smaller ``silent" spectrograms dataset. Adversarial issues aside, these techniques have the potential to alleviate the need for an interactive IVR Turing test in production.

\section{Acknowledgement}
\label{sec:acknowledgement}

The authors thank the following Marchex colleagues for their efforts on data collection and system deployment: Dave Olszewski, Ryan O'Rourke and Shawn Wade.

\bibliographystyle{abbrv}
\bibliography{refs}
\end{document}